# Improved CNN Prediction Based Reversible Data Hiding

Yingqiang Qiu, Wanli Peng, Xiaodan Lin, Huanqiang Zeng, *Senior Member, IEEE*, and Zhenxing Qian, *Member, IEEE*

*Abstract*—This letter proposes an improved CNN predictor (ICNNP) for reversible data hiding (RDH) in images, which consists of a feature extraction module, a pixel prediction module, and a complexity prediction module. Due to predicting the complexity of each pixel with the ICNNP during the embedding process, the proposed method can achieve superior performance than the CNN predictor-based method. Specifically, an input image does be first split into two different sub-images, i.e., the "Dot" image and the "Cross" image. Meanwhile, each sub-image is applied to predict another one. Then, the prediction errors of pixels are sorted with the predicted pixel complexities. In light of this, some sorted prediction errors with less complexity are selected to be efficiently used for low-distortion data embedding with a traditional histogram shift scheme. Experimental results demonstrate that the proposed method can achieve better embedding performance than that of the CNN predictor with the same histogram shifting strategy.

## I. INTRODUCTION

REVERSIBLE data hiding (RDH) can losslessly recover both the embedded data and the cover medium [1]. Due to the trait, RDH has gradually become a hot research field in the information hiding community and has been widely used in several realistic scenarios [1], including medical, military, and law forensics et. al. According to the domain hiding a secret message, RDH can be categorized as two main branches: spatial domain-based RDH [2-23] and JPEG domain-based RDH [24-27]. The spatial domain-based RDH generally exploits three technologies, i.e., lossless compression (LC) [2-4], difference expansion (DE) [5-15], and histogram shifting (HS) [16-23]. While the JPEG domain-based RDH is mainly based on DCT coefficients modification [24, 25] or Huffman table modification [26, 27].

Currently, in the RDH community, pixel prediction has become a critical problem, which dramatically affects the performance of RDH algorithms [14]. The traditional predictors include the median edge direction (MED) predictor [6], interpolation predictor [7], gradient-adjusted predictor (GAP) [8], pixel-value-ordering (PVO) predictor [9, 12, 22], linear predictor [10], rhombus predictor [17-20], and ridge regression predictor [23], etc. Although these predictors have achieved supervising improvement, there is still a notable weak point, that is few neighboring pixels are used for pixel prediction [14]. If more adjacent pixels are served as reference pixels, higher prediction performance can be achieved. Due to its strong capabilities of different receptive fields fusion and whole optimization, a convolutional neural network (CNN) can be established and trained to predict pixels accurately by building a non-linear mapping for pixel prediction. In light of this, Luo et al. [13] presented a CNN-based stereo image RDH method by leveraging the correlations between the left view and the right view. Hu et al. [14] proposed a CNN predictor (CNNP) based RDH method, where a grayscale image was split into two sub-images, and each one is predicted with another one alternatively by using the CNNP. After that, Hu et al. [15] divided an image into four parts, and each part was predicted with the other three parts in turn by using a CNNP for a better prediction performance. In addition, a better visual quality of the marked image is achieved through adaptive embedding. Overall, the prediction performance of CNN predictors can be better than that of the traditional predictors.

From the above discussion, in order to improve performance, the existing methods conduct pixel prediction by leveraging adjacent pixels. While these methods [13-15] ignore the complexity of each pixel with deep learning, which limits the performance of RDH.

To tackle the above limitation, in this letter, we improve the CNNP presented in [14] by adding a complexity prediction part to predict the pixel's complexities precisely, which is called improved CNNP (ICNNP) in the rest of this letter. Specifically, during data embedding, we first split a grayscale image into two sub-images, where one sub-image is predicted by other one. Then, we sort the prediction errors of the predicted pixels according to their complexities, and the prediction errors with less complexity are used for data embedding with a classical HS

This work was supported in part by the National Key R&D Program of China (Grant 2021YFE0205400), the Natural Science Foundation of Xiamen, China (Grant 3502Z20227192), and the Natural Science Foundation of China (Grant U20B2051, 61972168, 62072114, 62002124, 61871434). Corresponding author: Zhenxing Qian.

Y. Qiu, X. Lin, and H. Zeng are with the College of Information Science & Engineering, Huaqiao University, Xiamen 361021, China. (e-mail: yqqiu@hqu.edu.cn, echo.linxd@gmail.com, zeng0043@hqu.edu.cn).
W. Peng and Z. Qian are with the School of Computer Science, Fudan University, Shanghai 200433, China. (e-mail: pengwanli@fudan.edu.cn, zxqian@fudan.edu.cn).
Y. Qiu and W. Peng contribute equally to this work.

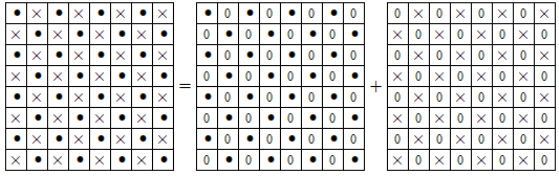

(a) Original image $I$  (b) "Dot" image $I_1$  (c) "Cross" image $I_2$

Fig. 1. Illustration of splitting an original image into two sub-images.

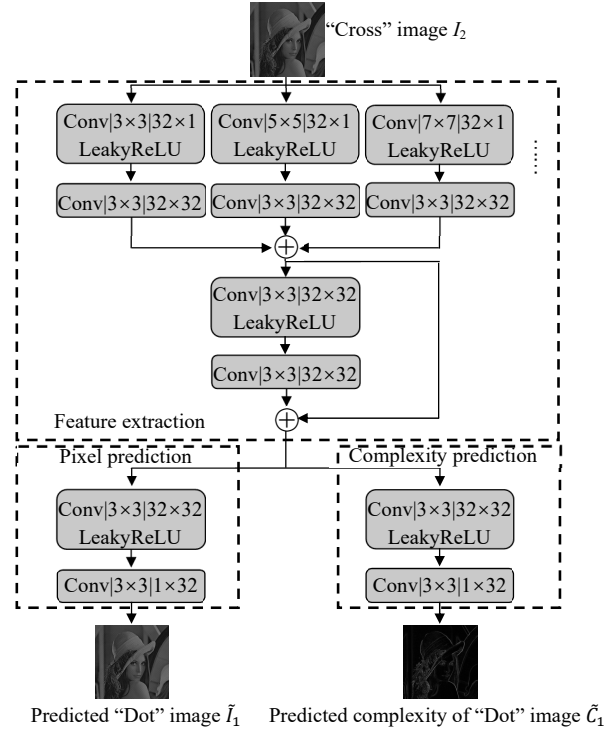

Fig. 2. The overall architecture of the proposed ICNNP.

strategy. Finally, experimental results show that the performance of the proposed method is better than that of the CNNP presented in [14].

The rest of this letter is organized as follows. The proposed improved RDH method is described in detail in Section II, and the experimental results and discussions are provided in Section III. Finally, we conclude our work in Section IV.

## II. PROPOSED IMPROVED METHOD

### A. Network Architecture

As shown in Fig. 1, according to the checkerboard context model [17], the original image is split into two sub-images which consist of "Dot" and "Cross" pixels, respectively. For the "Dot" image, the values of the "Dot" pixels are remained, while those of the "Cross" pixels are set to 0. Meanwhile, just the values of the "Dot" pixels are set to 0 for the "Cross" image. Based on the pixel correlation of the two sub-images, each sub-image is applied to predict the pixel values and complexities of another sub-image.

The overall of the proposed ICNNP is shown in Fig. 2. The architecture of the ICNNP is composed of three parts, i.e., feature extraction, pixel prediction, and complexity prediction. The "Cross" image $I_2$ is fed to the network to predict the values and complexities of the "Dot" pixels, where the values of complexity are adjusted to [0,255] for a good visualization display. The lower value, the lower complexity. The feature extraction consists of some convolution layers with different filter sizes (3 × 3, 5 × 5, 7 × 7, ⋯), which are parallelized and appended with a 3×3 convolution layer respectively to extract features from different receptive fields. A residual block is then applied to further aggregate and refine the learnt features from different branches. With the extracted feature, the pixel prediction yields the predicted "Dot" image $\tilde{I}_1$, and the complexity prediction yields the predicted complexity $\tilde{C}_1$ of the "Dot" image $I_1$. "Conv" stands for the convolution unit with kernel size $S × S$ and the number of channels is *output × input*. A LeakyReLU activation function [28] locates between each two convolution layers.

It is worthy to note that the complexity prediction is similar to the pixel prediction, i.e., instead of orthogonal adjacent pixels [14, 17], more adjacent pixels are used to nonlinearly predict the complexity of the pixel area, improving the performance of RDH.

### B. Training

In the ICNNP, the well-trained parameters of CNNP [14] are loaded into the feature extraction and pixel prediction. Note that, these parameters are fixed and the parameters of complexity prediction are updated during the training of the ICNNP. In the training, the input is the "Cross" image $I_2$, the outputs are the predicted "Dot" image $\tilde{I}_1$ and the predicted complexity $\tilde{C}_1$ of the "Dot" image $I_1$. Since the filter parameters of the feature extraction and the pixel prediction are fixed, the target is no longer the "Dot" image $I_1$ but the referenced complexity $C_1$ of $I_1$. The definition of $C_1$ is described as follows.

(1) For "Cross" pixels, $C_1(i,j)$ is set to 0;

(2) For "Dot" pixels, if $i = 1$ or $i = M$ or $j = 1$ or $j = N$, $C_1(i,j)$ is set to 0; otherwise, $C_1(i,j)(2 \leq i \leq M-1, 2 \leq j \leq N-1)$ is calculated as

$$C_1(i,j) = \frac{1}{R} \cdot \sqrt{\sum_{k=k_1}^{k_2}\sum_{l=l_1}^{l_2}(I(i+k,j+l) - I(i,j))^2}, \quad (1)$$

where,

$$\begin{cases} k_1 = -1, k_2 = 2 & ,i = 2 \\ k_1 = -2, k_2 = 1 & ,i = M-1 \\ k_1 = -2, k_2 = -2, 2 < i < M-1 \end{cases}, \quad (2)$$

$$\begin{cases} l_1 = -1, l_2 = 2 & ,j = 2 \\ l_1 = -2, l_2 = 1 & ,j = N-1 \\ l_1 = -2, l_2 = -2, 2 < j < N-1 \end{cases}, \quad (3)$$

$$R = (k_2 - k_1 + 1) \times (l_2 - l_1 + 1) - 1. \quad (4)$$

In the proposed method, the max predicted pixel area is 5 × 5, which can accurately calculate the pixel complexity.

As with CNNP in [14], we leveraged back-propagation [29] and Adam algorithm [30] to optimize the objective function defined as below:

$$\text{Loss} = \frac{1}{N}\sum_{i=1}^{N}(\tilde{C}_1 - C_1)^2 + \lambda\|\omega\|_2^2, \quad (5)$$

where $N$ stands for the number of training examples, $\omega$ represents all weights of the network, and $\lambda$ denotes the weight decay.

### C. ICNNP based RDH

Fig. 3 depicts the data embedding architecture of the ICNNP-based RDH method. The adopted double embedding strategy [17] with the HS technique [6] involves consecutive usage of the "Dot" embedding and the "Cross" embedding, and the "Cross" embedding is achieved after the "Dot" embedding.

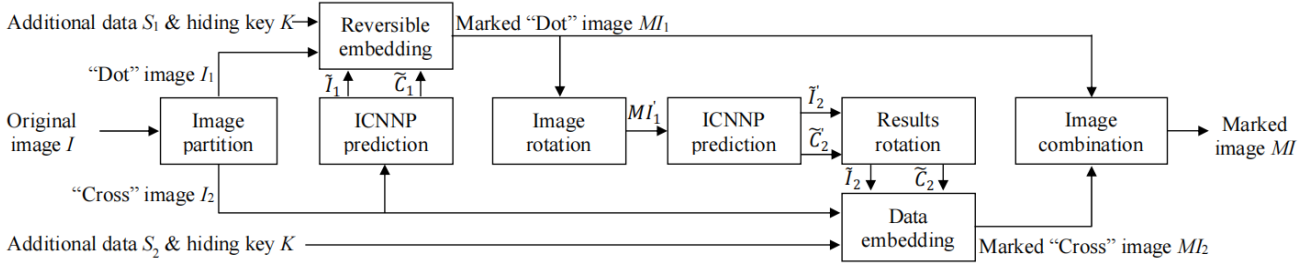

Fig. 3. The flowchart of data embedding by using the ICNNP based RDH method.

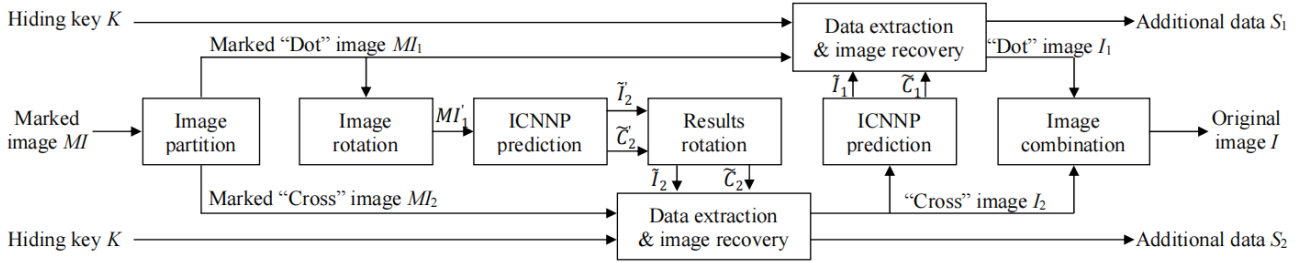

Fig. 5. The flowchart of data extraction and image recovery by using the ICNNP based RDH method.

The original image $I$ is firstly divided into two sub-images, i.e., a "Dot" image $I_1$ and a "Cross" image $I_2$. Next, the predicted "Dot" image $\tilde{I}_1$ and the predicted complexity $\tilde{C}_1$ of $I_1$ are predicted with $I_2$ as follows:

$$\{\tilde{I}_1, \tilde{C}_1\} = ICNNP(I_2). \qquad (6)$$

Then, the prediction errors of $I_1$ are calculated as

$$e_1(i,j) = I_1(i,j) - \tilde{I}_1(i,j), (i+j) \bmod 2 \equiv 0. \qquad (7)$$

According to the magnitude of predicted complexities and the size of the additional data $S_1$, we select the predicted errors with less complexity and determine two threshold $T_{n1}$ ($T_{n1}<0$) and $T_{p1}$ ($T_{p1} \geq 0$) for HS-based data embedding, which is achieved as

$$E_1(i,j) = \begin{cases} 2e_1(i,j) + b & ,if\ e_1(i,j) \in [T_{n1}, T_{p1}] \\ e_1(i,j) + T_{p1} + 1 & ,if\ e_1(i,j) > T_{p1} \\ e_1(i,j) + T_{n1} & ,if\ e_1(i,j) < T_{n1} \end{cases}, \qquad (8)$$

where $b \in [0,1]$ is the embedded data including the encrypted additional data and some auxiliary data [17]. Therefore, the marked "Dot" image $MI_1$ is generated as

$$MI_1(i,j) = \tilde{I}_1(i,j) + E_1(i,j). \qquad (9)$$

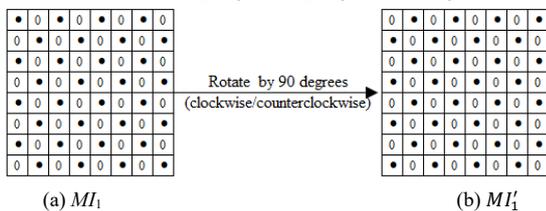

(a) $MI_1$          (b) $MI'_1$

Fig. 4. Illustration of image rotation of marked "Dot" image.

During the "Cross" embedding, due to the checkerboard pattern, the marked "Dot" image $MI_1$ cannot be fed into the network to predict the "Cross" image $I_2$ directly. As illustrated in Fig. 4, if the height/width is even, after being rotated clockwise/counterclockwise by 90 degrees, the resulting image $MI'_1$ has the same pattern as the "Cross" image $I_2$. With the same ICNNP, as shown in Eqn. (10), we feed the rotated marked "Dot" image $MI'_1$ into the network, then we obtain the predicted rotated "Cross" image $\tilde{I}'_2$ and its complexity $\tilde{C}'_2$.

$$\{\tilde{I}'_2, \tilde{C}'_2\} = ICNNP(MI'_1). \qquad (10)$$

Then, $\tilde{I}'_2$ and $\tilde{C}'_2$ are rotated counterclockwise/clockwise by 90 degrees to get the predicted "Cross" image $\tilde{I}_2$ and the predicted complexity $\tilde{C}_2$ of $I_2$. Similar to the "Dot" embedding, another part of additional data $S_2$ is encrypted with the hiding key $K$ and embedded into $I_2$ to obtain the marked "Cross" image $MI_2$.

Finally, we combine the marked "Dot" image $MI_1$ and the marked "Cross" image $MI_2$ to obtain the marked image $MI$.

Fig. 5 describes the architecture of data extraction/image recovery. Data extraction and image recovery are the inverse procedures of data embedding, so we operate the "Dot" extraction/recovery ahead of the "Cross" extraction/recovery. The marked image $MI$ is firstly divided into two sub-images, i.e., the marked "Dot" image $MI_1$ and the marked "Cross" image $MI_2$. With the rotated marked "Dot" image $MI'_1$, $\tilde{I}'_2$ and $\tilde{C}'_2$ are predicted by using the ICNNP as Eqn. (10). $\tilde{I}_2$ and $\tilde{C}_2$ are then obtained by rotating $\tilde{I}'_2$ and $\tilde{C}'_2$ respectively. Next, the marked prediction errors of $I_2$ are calculated as

$$E_2(i,j) = MI_2(i,j) - \tilde{I}_2(i,j), (i+j) \bmod 2 \equiv 1. \qquad (11)$$

According to the sorted magnitude of $\tilde{C}_2$ and the extracted auxiliary data $T_{n2}$ ($T_{n2}<0$) and $T_{p2}$ ($T_{p2} \geq 0$) as threshold, the embedded data is extracted as

$$b = E_2(i,j) \bmod 2, \ E_2(i,j) \in [2T_{n2}, 2T_{p2}+1], \qquad (12)$$

and the original prediction errors of $I_2$ are recovered as

$$e_2(i,j) = \begin{cases} \lfloor E_2(i,j)/2 \rfloor & ,if\ E_2(i,j) \in [2T_{n2}, 2T_{p2}+1] \\ E_2(i,j) - T_{p2} - 1 & ,if\ E_2(i,j) > 2T_{p2}+1 \\ E_2(i,j) - T_{n2} & ,if\ E_2(i,j) < 2T_{n2} \end{cases}, \qquad (13)$$

where $\lfloor \cdot \rfloor$ is the floor function. We decrypt the extracted bits to get $S_1$ with the hiding key $K$, and recover the original "Cross" image $I_2$ as

$$I_2(i,j) = e_2(i,j) + \tilde{I}_2(i,j), (i+j) \bmod 2 \equiv 1. \qquad (14)$$

Similarly, the embedded data $S_1$ is extracted correctly and the original "Dot" image $I_1$ is recovered losslessly. Finally, we combine the recovered "Cross" image $I_2$ and "Dot" image $I_1$ to obtain the original image $I$.

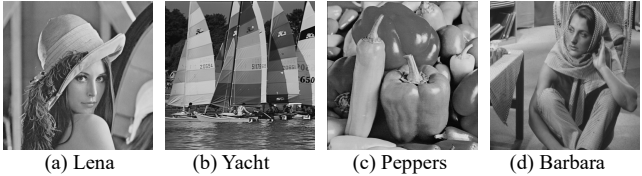

(a) Lena     (b) Yacht     (c) Peppers     (d) Barbara

Fig. 6. Four cover images.

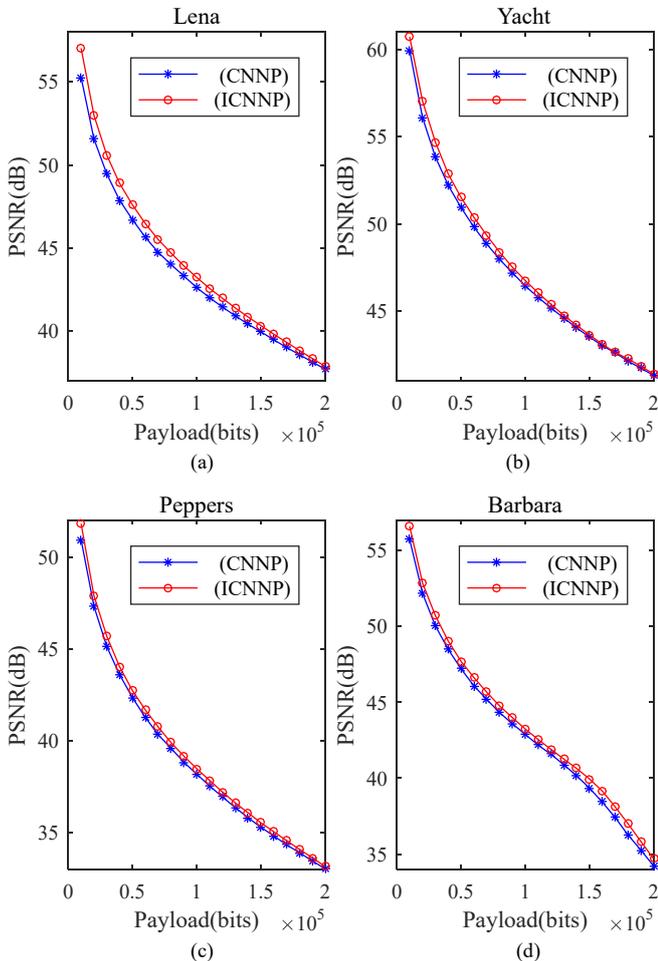

Fig. 7. Performance comparison of CNNP in [14] and the proposed ICNNP for RDH in four benchmark images.

## III. EXPERIMENTAL RESULTS

To evaluate the efficiency of the proposed ICNNP, the parameters of its complexity prediction model are trained by using 1,000 images randomly selected from BOWS-2 [31]. The proposed ICNNP is trained on an Intel Core i10 CPU (3.6 GHz) with 16 GB RAM and NVIDIA GeForce RTX 2060. The weight decay $\lambda$ is set to $1\times10^{-3}$, the batch size is 4 and the initial learning rate is $1\times10^{-3}$. In [14], the prediction accuracy of CNNP is proved better than some traditional linear predictors, including the BIP, MEDP, GAP, and DP, and the achieved rate-distortion performance is better than that of the above predictor with the expansion embedding scheme, it's also better than that of BIP with the HS scheme, and the performance of HS is far better than that of expansion embedding with the CNNP.

Therefore, we justly evaluate the performance with ICNNP by comparing it with that of CNNP [14] with the same HS technique.

With four benchmark images in Fig. 6 as cover images at the same embedding rate, we employ the peak signal-to-noise ratio (PSNR) between the original image and the marked image as the metric for objective image quality evaluation. Besides, we randomly select 100 images different from the training images from BOWS-2 [31] and test them with different embedding capacities to evaluate the universality of ICNNP.

Fig. 7 shows the PSNR values of four test images (Lena, Yacht, Peppers, and Barbara) with embedding capacities from 10,000 to 200,000 bits. From this figure, we can see that the PSNR values of the RDH method with the ICNNP are larger than those of the CNNP-based RDH method. Moreover, Table I shows the average PSNR values of the 100 test images for different embedding capacities. When the embedding capacity is 10,000 bits, the average PSNR value of the proposed ICNNP based RDH method is 62.37 dB, which is 1.06dB higher than that of the CNNP-based RDH method. Along with the embedding capacity increases from 20,000 to 150,000 bits, the average PSNR values of the improved RDH method are still 0.81dB, 0.60dB, 0.55dB, 0.56dB, 0.55dB, 0.54dB, 0.53dB, 0.51dB, 0.47dB, 0.41dB, 0.37dB, 0.33dB, 0.26dB, and 0.20dB higher respectively.

TABLE I
AVERAGE PSNR (dB) OF 100 IMAGES OF THE PROPOSED ICNNP-BASED METHOD AND THE CNNP-BASED METHOD [14]

| Embedding Capacity (bits) | CNNP | ICNNP |
|---|---|---|
| 10000 | 61.31 | **62.37** |
| 20000 | 58.01 | **58.82** |
| 30000 | 55.98 | **56.58** |
| 40000 | 54.43 | **54.98** |
| 50000 | 53.10 | **53.66** |
| 60000 | 51.94 | **52.49** |
| 70000 | 50.86 | **51.40** |
| 80000 | 49.85 | **50.38** |
| 90000 | 48.88 | **49.39** |
| 100000 | 47.91 | **48.38** |
| 110000 | 46.95 | **47.36** |
| 120000 | 46.03 | **46.40** |
| 130000 | 45.13 | **45.46** |
| 140000 | 44.29 | **44.55** |
| 150000 | 43.47 | **43.67** |

## IV. CONCLUSION

In this letter, we propose an improved CNN predictor for RDH, which extracts features from different receptive fields with whole optimization and uses more neighboring pixels to precisely predict the pixel value and its complexity. During data embedding, a grayscale image is split into two sub-images, one sub-image is applied to predict another sub-image alternately by using the ICNNP. Then the pixels' prediction errors are sorted according to the predicted pixels' complexities, and the prediction errors with less complexity are selected for data embedding with the classical HS strategy. The original image is recovered losslessly after the embedded data is extracted correctly, and the data extraction and image recovery are separable. Experimental results show that the achieved performance of the improved CNNP with the classical HS strategy is better than that of the CNNP presented in [14] with the same HS strategy.

## References

[1] Y. -Q. Shi, X. Li, X. Zhang, H. Wu, and B. Ma, "Reversible data hiding: Advances in the past two decades," *IEEE Access*, vol. 4, pp. 3210–3237, 2016.

[2] M. U. Celik, G. Sharma, A. M. Tekalp, and E. Saber, "Lossless generalized-LSB data embedding," *IEEE Trans. Image Process.*, vol.14, no. 2, pp. 253–266, 2005.

[3] W. Zhang, X. Hu, X. Li, and N. Yu, "Optimal transition probability of reversible data hiding for general distortion metrics and its applications," *IEEE Trans. Image Process.*, vol. 24, no. 1, pp. 294–304, 2015.

[4] D. Hou, W. Zhang, Y. Yang, and N. Yu, "Reversible data hiding under inconsistent distortion metrics," *IEEE Trans. Image Process.*, vol. 27, no. 10, pp. 5087–5099, Oct. 2018.

[5] J. Tian, "Reversible data embedding using a difference expansion," *IEEE Trans. Circuits Syst. Video Technol.*, vol. 13, no. 8, pp. 890–896, Aug. 2003.

[6] D. M. Thodi and J. J. Rodriguez, "Expansion embedding techniques for reversible watermarking," *IEEE Trans. Image Process.*, vol. 16, no. 3, pp. 721–730, Mar. 2007.

[7] L. Luo, Z. Chen, M. Chen, X. Zeng, and Z. Xiong, "Reversible image watermarking using interpolation technique," *IEEE Trans. Inf. Forensics Secur.*, vol. 5, no. 1, pp. 187–193, Mar. 2010.

[8] D. Coltuc, "Low distortion transform for reversible watermarking," *IEEE Trans. Image Process.*, vol. 21, no. 1, pp. 412-417, Jan. 2012.

[9] X. Li, J. Li, B. Li, and B. Yang, "High-fidelity reversible data hiding scheme based on pixel-value-ordering and prediction-error expansion," *Signal Process.*, vol. 93, no. 1, pp. 198–205, 2013.

[10] I.-C. Dragoi and D. Coltuc, "Local-prediction-based difference expansion reversible watermarking," *IEEE Trans. Image Process.*, vol. 23, no. 4, pp. 1779–1790, Apr. 2014.

[11] Y. Qiu, Z. Qian, and L. Yu, "Adaptive reversible data hiding by extending the generalized integer transformation," *IEEE Signal Process. Lett.*, vol. 23, no. 1, pp. 130–134, Jan. 2016.

[12] W. He and Z. Cai, "An insight into pixel value ordering prediction-based prediction-error expansion," *IEEE Trans. Circuits Syst. Video Technol.*, vol. 15, pp. 3859-3871, 2020.

[13] T. Luo, G. Jiang, M. Yu, C. Zhong, H. Xu, and Z. Pan, Convolutional neural networks-based stereo image reversible data hiding method, *J. Vis. Commun. Image Represent.*, vol. 61, pp. 61–73, 2019.

[14] R. Hu and S. Xiang, "CNN Prediction Based Reversible Data Hiding," *IEEE Signal Process. Lett.*, vol. 28, pp. 464-468, 2021.

[15] R. Hu and S. Xiang, "Reversible Data Hiding by Using CNN Prediction and Adaptive Embedding," *IEEE Trans. Pattern Anal. Mach. Intell.*, to be published.

[16] Z. Ni, Y. Shi, N. Ansari, and W. Su, "Reversible data hiding," *IEEE Trans. Circuits Syst. Video Technol.*, vol. 16, no. 3, pp. 354–362, Mar. 2006.

[17] V. Sachnev, H. J. Kim, J. Nam, S. Suresh, and Y. -Q. Shi, "Reversible watermarking algorithm using sorting and prediction," *IEEE Trans. Circuits Syst. Video Technol.*, vol. 19, no. 7, pp. 989–999, 2009.

[18] X. Li, W. Zhang, X. Gui, and B. Yang, "Efficient reversible data hiding based on multiple histograms modification," *IEEE Trans. Inf. Forensics Secur.*, vol. 10, no. 9, pp. 2016–2027, Sep. 2015.

[19] W. Qi, X. Li, T. Zhang, and Z. Guo, "Optimal Reversible Data Hiding Scheme Based on Multiple Histograms Modification," *IEEE Trans. Circuits Syst. Video Technol.*, vol. 30, no. 8, pp. 2300-2312, Aug. 2020.

[20] J. Wang, X. Chen, J. Ni, N. Mao, and Y. Shi, "Multiple Histograms-Based Reversible Data Hiding: Framework and Realization," *IEEE Trans. Circuits Syst. Video Technol.*, vol. 30, no. 8, pp. 2313-2328, Aug. 2020.

[21] B. Ou and Y. Zhao, "High capacity reversible data hiding based on multiple histograms modification," *IEEE Trans. Circuits Syst. Video Technol.*, vol. 30, no. 8, pp. 2329-2342, Aug. 2020.

[22] T. Zhang, X. Li, W. Qi, and Z. Guo, "Location-based pvo and adaptive pairwise modification for efficient reversible data hiding," *IEEE Trans. Inf. Forensics Secur.*, vol. 15, pp. 2306–2319, 2020.

[23] X. Wang, X. Wang, B. Ma, Q. Li, and Y. -Q. Shi, "High Precision Error Prediction Algorithm Based on Ridge Regression Predictor for Reversible Data Hiding," *IEEE Signal Process. Lett.*, vol. 28, pp. 1125-1129, 2021.

[24] F. Huang, X. Qu, H. J. Kim, and J. Huang, "Reversible data hiding in JPEG images," *IEEE Trans. Circuits Syst. Video Technol.*, vol. 26, no. 9, pp. 1610–1621, Sep. 2016.

[25] D. Hou, H. Wang, W. Zhang, and N. Yu, "Reversible data hiding in JPEG image based on DCT frequency and block selection," *Signal Process.*, vol. 148, pp. 41–47, 2018.

[26] Y. Qiu, Z. Qian, H. He, H. Tian, and X. Zhang, "Optimized lossless data hiding in JPEG bitstream and relay transfer based extension," *IEEE Trans. Circuits Syst. Video Technol.*, vol. 31, no. 4, pp. 1380-1394, Apr. 2021.

[27] Y. Du, Z. Yin, and X. Zhang, "High capacity lossless data hiding in JPEG bitstream based on general VLC mapping," *IEEE Trans. Depend. Secure Comput.*, vol. 19, no. 2, pp. 1420-1433, 2022.

[28] A. L. Maas, Y. H. Awni, and Y. N. Andrew, Rectifier nonlinearities improve neural network acoustic models. in *Proceedings of the 30th International Conference on Machine Learning* (ICML-13), 2013.

[29] Y. Lecun, L. Bottou, Y. Bengio, and P. Haffner, "Gradient-based learning applied to document recognition," in *Proceedings of the IEEE*, vol. 86, no. 11, pp. 2278-2324, Nov. 1998.

[30] D. Kingma and J. Ba, "Adam: A method for stochastic optimization," in *Proc. Int. Conf. Learn. Representations*, 2015.

[31] P. Bas and T. Furon, "Image database of bows-2," 2017. [Online]. Available at http://bows2.ec-lille.fr/.